\documentclass[a4paper,twocolumn]{article}
\usepackage[utf8]{inputenc}
\usepackage[UKenglish]{babel}
\usepackage{tgpagella}
\usepackage{amsmath,amssymb,amsthm}
\usepackage{url}
\usepackage{cleveref}
\usepackage{mathpartir}
\usepackage{mathtools}
\usepackage{xfrac}
\usepackage[disable]{todonotes}

\usepackage[margin=1.8cm,top=1.5cm]{geometry}
\newlength{\bibitemsep}
\newlength{\bibparskip}
\let\oldthebibliography\thebibliography
\renewcommand\thebibliography[1]{%
  \oldthebibliography{#1}%
  \setlength{\parskip}{\bibitemsep}%
  \setlength{\itemsep}{\bibparskip}%
}

\usepackage{titlesec}
\titleformat*{\section}{\large\bfseries}

\theoremstyle{definition}
\newtheorem{theorem}{Theorem}
\newtheorem{proposition}[theorem]{Proposition}

\newtheorem{definition}[theorem]{Definition}

\newtheorem{example}[theorem]{Example}

\newcommand{\cn}{\textsuperscript{[citation needed]}}

\newcommand{\bN}{\mathbb N}

\newcommand{\bR}{\mathbb R}
\newcommand{\sets}{\mathsf{Sets}}

\newcommand{\tprob}{{T_\mathsf{prob}}}
\newcommand{\tcoin}{{T_\mathsf{coin}}}

\newcommand{\hide}[1]{}

\begin{document}
\title{\vspace{-0.5cm}Denotational semantics for languages for inference: \\semirings, monads, and tensors\vspace{-0.2cm}}
\author{Cristina Matache \\ \small University of Edinburgh \\ \small cristina.matache@ed.ac.uk \and Sean Moss \\ \small University of Oxford \\ \small sean.moss@cs.ox.ac.uk \and Sam Staton \\ \small University of Oxford \\ \small sam.staton@cs.ox.ac.uk \and Ariadne Si Suo \\ \small University of Oxford \\ \small si.suo@sjc.ox.ac.uk}
\date{}
\maketitle
\section{Overview}
\newcommand{\srIO}[1]{R^{\mathsf{io}}_{#1}}
\newcommand{\srSt}[1]{R^{\mathsf{st}}_{#1}}
\newcommand{\tIO}[1]{T^{\mathsf{io}}_{#1}}
\textbf{Semirings} abound both in program analysis (e.g.~\cite{kozen-kat,probnetkat}) and in semantics for probabilistic programs (e.g.~\cite{lmmp-weighted-relational}). \textbf{Monads} abound in functional programming, in inference programming (e.g.~\cite{ap-selection}), and in probabilistic programming for statistical modelling, both in theory (e.g.~\cite{jlmz-commppl,jgl,vks-wqbs}) and in practice (e.g.~\cite{monad-bayes,ramsey-monad}). This talk is about the relationships between semirings and monads, and about new ways of building new semirings and new monads to interpret various different phenomena, and combinations of phenomena, in languages for inference. 

We summarize several examples of semirings:
\begin{itemize}
\item The non-negative reals, $\bR^+$, used for unnormalized measures in Bayesian statistics and other optimization scenarios;
\item The polynomials, $\bN[X,\bar X]$, thought of as unnormalized probabilities with an unknown Bernoulli parameter~$X$ (see~\Cref{example:tcoin});
\item For input/output messages, we consider a semiring~$\srIO I$ freely generated by $\{\mathsf{input}_i~|~i\in I\}\cup\{\mathsf{output}_i~|~i\in I\}$, for some finite set of input/output messages $I$ (see~\Cref{example:tio}).
\item For state, we consider a semiring~$\srSt I$ generated by $\{\mathsf{read}_i~|~i\in I\}\cup\{\mathsf{write}_i~|~i\in I\}$, for some finite set of storable values $I$ (see Appendix~A for details).
\end{itemize}
The key methods we propose are \emph{tensors} and \emph{convexity classes}.
\paragraph{Tensors and commutativity.} We show that the tensor of semirings corresponds to the tensor of monads, as studied in~\cite{hyland-plotkin-power-combining-computational-effects}. Thus the tensor of semirings is a method for combining monads for programming. Moreover, the tensor of semirings corresponds to the idea that in the combined monad, the lines of a program can be reordered (commute), i.e.
\[
\texttt{x <- t ; y <- u}  \quad =\quad  \texttt{y <- u ; x <- t}
\]
where \verb|t| and \verb|u| come from monads for different semirings.
This reordering of lines is important as a program optimization, particularly in probabilistic programming~\cite{mlkbs-rao-blackwell,sam-sfinite}, and moreover connects intuitively to conditional independence of \verb|x| and \verb|y|.

\paragraph{Monads via convexity classes.} Many semirings come with a particular subset which we call a `convexity class'. For example, the semirings $\bR^+$ and $\bN[X,\bar X]$ have convexity class $\{1\}$, and the semirings for state and I/O have convexity classes too (see \Cref{example:tio} and Appendix~A). Any convexity class in a semiring induces a monad. We show that the probability, state and I/O monads all arise in this way, giving a new `convex' understanding of state and I/O. The semiring $\bN[X,\bar X]$ induces a new monad which is good for modelling programs with an unknown Bernoulli variable.

\medskip
 We then combine these two methods, tensors and convexity classes, to give methods for building models of phenomena across languages for inference. Specifically, we show that several tensors of monads can be computed via convexity classes in tensors of semirings.
For example:
$\bN[X,\bar X]\otimes \bR^+$ combines a Bernoulli variable with ordinary probability;
$\bN[X,\bar X]\otimes \bN[Y,\bar Y]$ models two unknown Bernoulli variables;
$\bR^+\otimes \srSt I$ could model a combination of probability and state (as in e.g.~\cite{afr-xrp});
$\bR^+\otimes \srIO I$ models the combination of probability and input-output, for an interactive probabilistic program (e.g. over streaming data~\cite{lw-streaming}).

\section{Algebraic theories and semirings}

Rather than general monads we work with finitary monads or \emph{algebraic theories}, which in practice we think of as \emph{abstract clones} (defined just like a Kleisli triple except $TX$ appears only for $X$ a \emph{finite} set).
For a theory $T$, we write $T(n)$ for the set of equality-classes of terms in $n$ variables, or equivalently free $T$-algebra on the set $\{x_1,\ldots,x_n\}$. This is the relationship between monads and algebraic effects (e.g.~\cite{plotkin-power-notions-of-computation-determine-monads}).

We omit precise definitions but roughly a semiring $R$ is a `ring without subtraction' and an $R$-module is an (additive) commutative monoid with a compatible (left) action of $R$.
Modules are the algebras of a theory $T_R$, where $T_R(n) \cong R^n$ identifying $(r_1,\ldots,r_n) \in R^n$ with $r_1x_1 + \ldots + r_nx_n$~\cite{elgot-matricial-theories,be-iteration}.
It is useful to think of $T_R(n)$ as the set of $R$-valued measures on $\{x_1,\ldots,x_n\}$.
Switching back to monads briefly, $T_R(X)$ is the set of finitely supported functions $X\to R$.
\begin{example}
  $T_{\bR^+}$ corresponds to the monad of finitely-supported \emph{unnormalized} measures.

\end{example}

\section{Convexity classes and theories}

\begin{definition}A subset $S$ of a semiring $R$ is a \emph{convexity class} if $1 \in S$ and whenever $a_1 + \ldots + a_n, b_1, \ldots, b_n \in S$ we have $a_1b_1 + \ldots + a_nb_n \in S$.
For such $(R,S)$, the theory of \emph{$(R,S)$-convex sets} is the subtheory $T_{R,S}$ of $T_R$ where the set $T_{R,S}(n)$ of $n$-ary operations consists of those $\sum_i r_ix_i \in T_R(n)$ where $\sum r_i \in S$.
We call such theories \emph{convexity theories}.
\end{definition}
In general, $T_{R,S}$ relates to $T_R$ as probability measures relate to unnormalized measures.

\begin{example}\label{example:convsets}
  The usual theory of convex sets~\cite{stone-barycentric} is $\tprob \coloneqq T_{\bR^+,\{1\}}$ and here $\tprob(n)$ is the set of probability measures on $\{x_1,\ldots,x_n\}$.
  Considering $T_{\bR^+,[0,1]}$ gives subprobability measures instead.

\end{example}

\begin{example}\label{example:tio}
  For any finite set $I$, let $\tIO I$ be the free theory with unary operations $\mathsf{output}_i$ for $i \in I$ and one $|I|$-ary operation $\mathsf{input}$ -- the I/O monad~\cite{moggi-notions-of-computation-and-monads}.
  Then $\tIO I$ is actually $T_{\srIO I,\Lambda_I}$ for the semiring $\srIO I$ from the introduction, for a suitable convexity class $\Lambda_I \subseteq \srIO I$.
  Concretely, elements of $\srIO I$ are finite multisets of lists in $\{\mathsf{input}_i, \mathsf{output}_i \mid i \in I\}$.

\end{example}

\begin{example}\label{example:tcoin}
  Let $\tcoin$ be the theory presented by a single binary operation $c$ satisfying the equations
  \begin{mathpar}
    c(x,x) = x
    \and
    c(c(w,x),c(y,z)) = c(c(w,y),c(x,z))
  \end{mathpar}
  i.e.\ $c$ is idempotent and commutes with itself.
  The operation $c$ represents a random binary choice made by a coin with an unknown bias: $c(x,y)$ means ``flip the coin and continue as $x$ if heads and as $y$ if tails''.
  Terms $t \in \tcoin(n)$ represent experimental procedures reporting one of $n$ outcomes according to a sequence of coin flips.
  With the aim of equating procedures which induce the same distribution on outcomes, we can discard irrelevant coin flips (idempotence) and swap the order of two consecutive coin flips (commutativity) \cite{ssyafr-beta-bernoulli}.

  Elaborating on some notation from the introduction, let $\bN[X,\bar X]$ stand for the semiring presented by $\langle X, \bar X \mid X + \bar X = 1, X \bar X = \bar X X \rangle$.
  This is like adding to $\bN$ an indeterminate real number $0 \leq p \leq 1$ together with $1-p$ (though we prefer notation that does not suggest subtraction).
  Indeed, one can show that the elements of $\bN[X,\bar X]$ are represented polynomials $P$ in $X,\bar X$ with coefficients in $\bN$ where $P = P'$ in $\bN[X,\bar X]$ iff $P[p/X,(1-p)/\bar X] = P'[p/X,(1-p)/\bar X]$ in $\bR^+$ for all real numbers $p \in [0,1]$.
  There is a theory homomorphism $\tcoin \to T_{\bN[X,\bar X]}$ induced by $c(x_1,x_2) \mapsto X x_1 + \bar X x_2$ which in general sends $t \in \tcoin(n)$ to $\sum_i P_i x_i$ where $P_i \in \bN[X,\bar X]$ is such that if the bias of the coin were actually $p \in [0,1]$ then $P_i[p/X,(1-p)/\bar X] \in [0,1]$ would be the actual probability of outcome $x_i$.
  We also have $\sum_i P_i = 1$, and in fact we get an isomorphism $\tcoin \cong T_{\bN[X,\bar X],\{1\}}$.
  Thus the semiring perspective confirms that the equational theory of $\tcoin$ equates experiments precisely when they induce the same distribution on outcomes for all possible values of the unknown bias $p$.
\end{example}

\section{Commutative combinations}
\label{sec:tensors}

The following standard construction \cite{hyland-plotkin-power-combining-computational-effects} lets one combine notions of computation so that they commute with each other.
The \emph{commutative combination} or \emph{tensor} $T_1 \otimes T_2$ of two theories is the universal theory admitting homomorphisms $\phi_1 : T_1 \to T_1 \otimes T_2$, $\phi_2 : T_2 \to T_1 \otimes T_2$ such that each term in the image of $\phi_1$ commutes with each term in the image of $\phi_2$.

The \emph{tensor} $R_1 \otimes R_2$ of two semirings is the universal semiring with homomorphisms $\phi_1 : R_1 \to R_1 \otimes R_2$, $\phi_2 : R_2 \to R_1 \otimes R_2$ such that $\phi_1(a) \cdot \phi_2(b) = \phi_2(b) \cdot \phi_1(a)$ for all $a \in R_1, b \in R_2$.

\begin{proposition}
  $T_{R_1} \otimes T_{R_2} \cong T_{R_1 \otimes R_2}$.
\end{proposition}
This lets us compute the theory tensor of two semiring theories, in terms of the usually more tractable semiring tensor.
The semiring tensor extends to semirings paired with convexity classes: $(R_1,S_1) \otimes (R_2,S_2) = (R_1 \otimes R_2, S_1 \otimes S_2)$ where $S_1 \otimes S_2$ is the smallest convexity class containing the images of $S_1$ and $S_2$.
This time we have a canonical comparison map
\begin{displaymath}
  \Phi_{(R,S),(R',S')} : T_{R,S} \otimes T_{R',S'} \to T_{R \otimes R', S \otimes S'}
\end{displaymath}
When $\Phi$ is an isomorphism, the tensor of theories can be understood in terms of the semiring tensor, as we now demonstrate.

\begin{example}
  For $p \in \bR^+$ let $\bN[p]$ be the subsemiring of $\bR^+$ generated by $p$.
  Then $\bN[\sfrac12] \otimes \bN[\sfrac13] \cong \bN[\sfrac16]$.
  Taking the convexity classes to be $\{1\}$, $\Phi$ happens in this case to give an isomorphism $T_{\bN[\sfrac12],\{1\}} \otimes T_{\bN[\sfrac13],\{1\}} \cong T_{\bN[\sfrac16],\{1\}}$.
  In programming terms, this says that a language with equiprobable binary and ternary choices is equivalent to one with a six-sided die.\todo{Actually write out a proof of this elsewhere.}
\end{example}

\begin{example}
  Terms of $\tprob \otimes \tcoin$ are simple generative models built from flipping a coin with unknown bias and sampling from known sources of randomness.
  The semiring $\bR^+[X,\bar X] \coloneqq \bR^+ \otimes \bN[X,\bar X]$ extends the polynomials of $\bN[X,\bar X]$ with coefficients in $\bR^+$, and admits a similar characterization of equality.
  The composite
  \begin{displaymath}
    \tprob \otimes \tcoin \cong T_{\bR^+,\{1\}}\otimes T_{\bN[X,\bar X],\{1\}} \xrightarrow{\Phi} T_{\bR^+[X,\bar X], \{1\}}
  \end{displaymath}
  sends $t \in (\tprob \otimes \tcoin)(n)$ to $\sum_i P_i x_i$ where $P_i \in \bR^+[X,\bar X]$ is a polynomial computing the probability of outcome $x_i$ (and $\sum_i P_i = 1$).
  In fact, the map is an isomorphism $\tprob \otimes \tcoin \cong T_{\bR^+[X,\bar X], \{1\}}$ (c.f.~\cite{ssyafr-beta-bernoulli}).
\end{example}

\begin{example}
  The map $\Phi$ is an isomorphism in the case of $(\bR^+,\{1\}) \otimes (\srIO I,\Lambda_I)$.
  Thus the combination of probability and I/O, $\tprob \otimes \tIO I$, is a convexity theory $T_{\bR^+ \otimes \srIO I, \{1\} \otimes \Lambda_I}$.
  Concretely, $\bR^+ \otimes \srIO I$ consists of finitely-supported functions from lists in $\{\mathsf{input}_i, \mathsf{output}_i \mid i \in I\}$ to $\bR^+$.
\end{example}

\section{Conclusion}

As we have shown, semirings give a fundamental perspective on many computational phenomena, and are also a tool to calculate the commutative combination of different effects with a focus on probability and inference.
Current work in progress is to find sufficient conditions for $\Phi$ to be an isomorphism.
In further work we hope to generalize away from finitary monads on $\sets$ to cover iteration (e.g.~\cite{be-iteration}) and more general classes of measures.

\bibliographystyle{abbrv}
\bibliography{lafi}
\appendix
\section{Appendix: Summary of semirings and convexity classes}
\newcommand{\strd}[1]{\mathsf{rd}_{#1}}
\newcommand{\stwr}[1]{\mathsf{wr}_{#1}}
A semiring is a set $R$ equipped with two associative binary operations $+$ and $\times$,
with units $0$ and $1$ respectively, such that $+$ is commutative ($r+s=s+r$) and
$\times$ distributes over $+$ on both sides.
\paragraph{Examples of semirings considered in this abstract:}
\begin{itemize}
\item The semiring $\bR^+$ comprises the positive reals with the usual additional and multiplication.
\item The polynomial semiring $\bN[X,\bar X]$ is generated by
  $\langle X, \bar X \mid X + \bar X = 1, X \bar X = \bar X X \rangle$. This means that it is the least semiring containing
  formal constants $X$ and $\bar X$ and additionally satisfying the given equations (Ex.~\ref{example:tcoin}).
\item The semiring for input/output $\srIO I$ is the free semiring generated by $\{\mathsf{input}_i~|~i\in I\}\cup\{\mathsf{output}_i~|~i\in I\}$. The inhabitants are formal sums of strings built from $\mathsf{input}_i$ and $\mathsf{output}_i$, where the multiplication is string concatenation
  (Ex.~\ref{example:tio}).
\item The semiring for state $\srSt I$ is generated by
  \[\hspace{-5mm}
\left \langle  \begin{array}{c} \strd i,\stwr i\\(i\in I)\end{array}~\left|~
  \begin{array}{l}
    \stwr i\strd i=\stwr i, \ \ 
  \stwr i \stwr j =\stwr j
    \\\stwr i  \strd j =0\,(i\neq j),\ \sum_{i\in I}\strd i \stwr i = 1
  \end{array}
  \right \rangle\right.
  \]
  (See~\cite{ak-skat},
  and compare with \cite{plotkin-power-notions-of-computation-determine-monads}.)
\item
  The tensor of semirings $R_1$ and $R_2$ is generated by
  \[\hspace{-5mm}
\left \langle  \!\begin{array}{c} \phi_1(r_1),\phi_2(r_2)\\(r_1{\in} R_1,r_2{\in} R_2)\end{array}\!\left|\!
  \begin{array}{l}
    \phi_1(r_1)\cdot \phi_2(r_2)=
    \phi_2(r_2)\cdot \phi_1(r_1)\\
    \phi_1,\phi_2\text{ are homomorphisms}
  \end{array}\!
  \right \rangle\right.
  \]
\end{itemize}
\paragraph{Convexity classes considered in this abstract:}
\begin{itemize}
\item We considered $\bR^+$ and $\bN[X,\bar X]$ with the singleton convexity class $\{1\}$.
\item We also considered $\bR^+$ with convexity class $[0,1]$, for subprobability (Ex.~\ref{example:convsets}).
\item For $\srIO I$, we built the convexity class as follows.
  A finite tree $t$ built from $I$-ary input nodes and unary output nodes can be regarded in terms of the set of
  paths~$p$ through the tree, which are given by strings from
  $\{\mathsf{input}_i~|~i\in I\}\cup\{\mathsf{output}_i~|~i\in I\}$.
  We consider the convexity class given by \[\textstyle \Lambda_I\coloneqq\{\sum_{p\in t} p~|~\text{$t$ is a valid tree}\}\subseteq \srIO I\text.\]
  In this way we recover the I/O monad~\cite{moggi-notions-of-computation-and-monads}.
  We remark that the individual paths, as strings, are reminiscent of traces in probabilistic programming (e.g.~\cite{lcscm-traces}).
\item
  For $\srSt I$, we consider the following convexity class. 
  A state transformer is a function $f:I\to I$.
  Any state transformer induces
  $(\sum_{i\in I} \strd i \stwr {f(i)})\in \srSt I$.
  We consider the convexity class
  \[
  \{\textstyle\sum_{i\in I} \strd i \stwr {f(i)}~|~f:I\to I\}\ \subseteq
  \srSt I\text.
  \]
  In this way we recover the global state monad from~\cite{moggi-notions-of-computation-and-monads,plotkin-power-notions-of-computation-determine-monads}.
\end{itemize}
We also considered tensors of convexity classes (\S\ref{sec:tensors}).  

\end{document}